# Asymptotic Performance of Coded Slotted ALOHA with Multi Packet Reception

Čedomir Stefanović, *Senior Member, IEEE,* Enrico Paolini, *Member, IEEE,* Gianluigi Liva, *Senior Member, IEEE*

*Abstract*—In this letter, we develop a converse bound on the asymptotic load threshold of coded slotted ALOHA (CSA) schemes with $K$-multi packet reception capabilities at the receiver. Density evolution is used to track the average probability of packet segment loss and an area matching condition is applied to obtain the converse. For any given CSA rate, the converse normalized to $K$ increases with $K$, which is in contrast with the results obtained so far for slotted ALOHA schemes based on successive interference cancellation. We show how the derived bound can be approached using spatially-coupled CSA.

## I. INTRODUCTION

The analysis of slotted random access protocols, like slotted ALOHA (SA) [1] or framed slotted ALOHA (FSA) [2], often assumes that collision slots are unusable. This assumption, inherited from [3], is the main factor limiting the achievable throughput to the well known bound of $1/e$ packet/slot. For framed schemes, this limitation was overcome in [4] via the use of successive interference cancellation (SIC). The contention resolution diversity slotted ALOHA (CRDSA) scheme [4] operates as follows: (i) users contend by transmitting packet replicas in several randomly chosen slots of the frame; (ii) decoding a packet from a singleton slot enables removal of its replicas from the slots in which they occurred via interference cancellation; (iii) slots from which replicas were removed may become singleton slots, triggering a new iteration. This SIC-based procedure yields a throughput of $0.55$ packet/slot for the case of two replicas per user.

A further step was the identification of key analogies with iterative erasure recovery algorithms used in modern channel codes, which enabled application of the coding-oriented tools to design and analyze SIC-based schemes [5]. Specifically, it was shown that irregular repetition slotted ALOHA (IRSA), where the number of replicas per user is selected according to an appropriate probability mass function (PMF), asymptotically achieves throughputs approaching $1$ packet/slot, which is the ultimate limit for the collision channel model.

Subsequent works applied various coding-oriented tools to the design of SIC-based slotted ALOHA schemes; we refer the reader to [6] for an overview. We particularly note coded slotted ALOHA (CSA) [7], in which users segment their packets, encode segments using segment-oriented linear block codes, and transmit encoded segments (instead of packet replicas). The advantage of CSA is that it can support combination of rates and loads which are not achievable by IRSA.

Previous work on CSA assumed contention on the collision channel model, where the receiver can correctly demodulate and decode any packet received in a singleton slot and no information can be extracted from a collided packet (unless the SIC cancels all contributions of interference impairing it). A generalization of this model is represented by multi packet reception (MPR) channels. In a basic MPR channel model, the receiver is assumed to be able to decode all packets in slots where no more than $K$ colliding packets are present and to extract no information from slots where more than $K$ packets are interfering each other. This model, hereafter referred to as $K$-MPR channel, appeared often in the literature, e.g., in [8]–[10] (note that the model reduces to the standard collision channel model for $K = 1$). An analysis of IRSA over the $K$-MPR channel was conducted in [10], where it was shown that, letting $\mathsf{G}^\star$ be the *load threshold* of the scheme (a quantity that corresponds to the number of innovative packets that can be successfully received in a slot and that is formally defined later), the ratio $\mathsf{G}^\star/K$ *decreases* with $K$ in most cases, for a fixed maximum number of replicas per contending user. Assuming that a $K$-MPR capability comes at the expense of a $K$-fold increase in the required time-frequency resources, cf. [10]–[13], this implies that investing in resources to achieve MPR capabilities is not fruitful in terms of $\mathsf{G}^\star/K$, although some benefits for finite frame lengths were reported [10].

In this letter, we make several advances regarding the previous work. First, we perform an asymptotic analysis of CSA over the $K$-MPR channel and establish a converse bound on the load threshold $\mathsf{G}^\star$ that is valid for *any* rate $0 < \mathsf{R} \leq 1$ (the rate represents the fraction of innovative data in the transmitted data, and is formally defined later). We then show that the bound normalized to $K$ *increases* with $K$, suggesting that a scheme performing close to it may benefit, in terms of $\mathsf{G}^\star/K$, from a higher MPR capability. Finally, we explicitly construct a spatially-coupled scheme approaching the bound and benefiting from a higher MPR capability.

The letter is organized as follows. Section II introduces the system model. Section III presents the asymptotic performance analysis of CSA, which is exploited in Section IV to derive the converse bound. In Section IV, an explicit scheme approaching the limit is also revealed. Section V concludes the letter.

Č. Stefanović is with Department of Electronic Systems, Aalborg University, 9220 Aalborg, Denmark, Email: cs@es.aau.dk.

E. Paolini is with CNIT, DEI, University of Bologna, via Venezia 52, 47521 Cesena (FC). E-mail: e.paolini@unibo.it.

G. Liva is with Institute of Communications and Navigation of the German Aerospace Center (DLR), Muenchner Strasse 20, 82234 Wessling, Germany, Email: Gianluigi.Liva@dlr.de.

The work of Č. Stefanovic was supported by the Danish Council for Independent Research, Grant No. DFF-4005-00281. The work of E. Paolini was supported by ESA/ESTEC under Contract No. 4000118331/16/UK/ND.

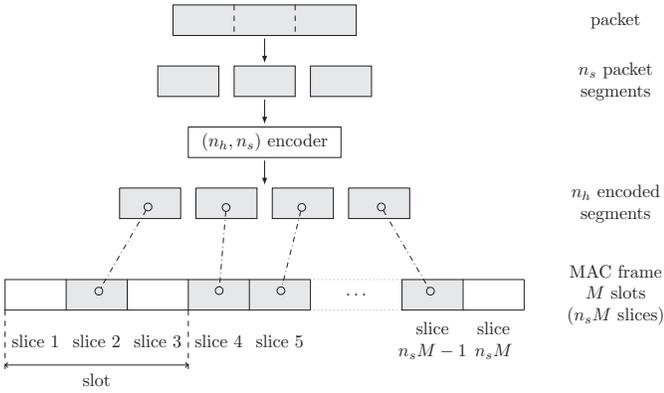

Fig. 1. Contention scheme from user perspective.

## II. SYSTEM MODEL

We adopt the system model from [7]. We conduct the analysis for a single batch arrival of $N$ users, each having a packet to transmit. The packets are of the same length, and the users contend to deliver their packets to a common receiver.

Prior to contention, each user divides its packet into $n_s$ equal-length segments and encodes them using a segment-level linear block $(n_h, n_s)$ code $C_h$. The obtained $n_h > n_s$ encoded segments (ESs) have the same length as the original segments; we refer to $n_h$ as the user degree. Code $C_h(n_h, n_s)$ is chosen independently at random by each user from a set of codes $\mathcal{C} = \{C_1, \ldots, C_\theta\}$, according to a PMF $\Lambda = \{\Lambda_1, \ldots, \Lambda_\theta\}$. The set $\mathcal{C}$ is known to the receiver. Each code $C_h$ has minimum distance of at least 2 and no idle symbols. The scheme reduces to IRSA when $\mathcal{C}$ only includes $(n_h, 1)$ repetition codes.

We focus on a single contention period, i.e., frame, which consists of $M$ slots of equal duration. Each slot is divided into $n_s$ slices, such that a slice accommodates transmission of one ES. All $N$ users are ready to transmit at the beginning of the frame and are frame- and slice-synchronized. A user selecting code $C_h$ contends by choosing uniformly randomly $n_h$ slices in the frame and by transmitting its $n_h$ ESs in these slices, as illustrated in Fig. 1. It is also assumed that each ES includes information about the code by which it was produced and about the slice location of the other $n_h - 1$ ESs (e.g., this information can be placed in the segment header).

We introduce the *rate* and the *load* of the scheme as

$$\mathsf{R} = \frac{n_s}{\bar{n}} \qquad \text{and} \qquad \mathsf{G} = \frac{N}{M} \tag{1}$$

respectively, where $\bar{n} = \sum_{h=1}^{\theta} \Lambda_h n_h$. The load represents the ratio between the number of transmitted packets (packet segments) and the number of slots (slices) in the frame. The *physical* load of the scheme, representing the number of ESs transmitted over the available slices, is $\mathsf{G}_0 = (\bar{n}\, N)/(n_s\, M) = \mathsf{G}/\mathsf{R}$. Contention is performed over the $K$-MPR channel, such that, if at most $K$ ESs collide in a slice, then all of them are perfectly decoded, while no ESs can be recovered from the slice otherwise.[1]

[1] In [11], [12] it was shown that $K$-MPR on Gaussian multiple access channel can be achieved using combination of compute-and-forward and coding for adder channels. Another, practical alternative is to use direct-sequence spreading, where $K$ corresponds to the spreading factor [13].

After the receiver has observed and stored all slices of a frame, it iteratively performs the following two-steps:
1) For each slice, if the number of the currently contained ESs in it does not exceed $K$, recover all ESs in the slice.
2) For each user packet of which at least one ES has been recovered, perform *maximum a posteriori* (MAP) erasure decoding of the corresponding code drawn from $\mathcal{C}$. For each new ES recovered through MAP erasure decoding, cancel its interference contribution from the slice where it was transmitted (thus reducing the number of ESs contained in the slice).

The iterations are run until no new ESs can be recovered. Interference cancellation is assumed ideal.

## III. ASYMPTOTIC ANALYSIS

We refer to the number of ESs per slice as the slice degree. In the limit where both the number $N$ of active users and the number $M$ of slots tend to infinity, with their ratio $\mathsf{G}$ being constant, the number of ESs per slice follows a Poisson distribution. Denoting by $P_i$ the probability that the slice degree is $i$ at the beginning of the SIC process, we have

$$P_i = \frac{\mathsf{G}_0^i}{i!} \exp\left(-\mathsf{G}_0\right), \; i \geq 0. \tag{2}$$

We proceed by analyzing the average probability that an ES remains unrecovered through the SIC procedure introduced in the previous section. We consider the asymptotic case, when $N \to \infty$, and base our approach on density evolution [14]. The evaluation tracks the average probabilities that an ES is *not* recovered after steps 1 and 2 of the each iteration.

Denote by $\mathsf{q}_\ell^{(h)}$ the probability that an ES remains unrecovered at the end of step 2 of iteration $\ell$, given that the ES was encoded by code $C_h$. Further, denote by $\mathsf{p}_\ell^{(i)}$ the probability that an ES remains unrecovered at the end of step 1 of iteration $\ell$, given that the current degree of the slice in which it was received is $i$. The average probability that an ES remains unrecovered at the end of step 2 of iteration $\ell$ is

$$\mathsf{q}_\ell = \sum_{h=1}^{\theta} \lambda_h \mathsf{q}_\ell^{(h)} \tag{3}$$

where $\lambda_h$ is the probability that an ES was transmitted by a user employing the code $C_h$, given by $\lambda_h = n_h \Lambda_h / \bar{n}$. Similarly, the average probability that an ES remains unrecovered at the end of step 1 of iteration $\ell$ is given by

$$\mathsf{p}_\ell = \sum_{i=1}^{\infty} \rho_i\, \mathsf{p}_\ell^{(i)} \tag{4}$$

where $\rho_i$ is the probability that an ES occurs in a slice of degree $i$, given by

$$\rho_i = \frac{\mathsf{G}_0^{i-1}}{(i-1)!} \exp\left(-\mathsf{G}_0\right), \; i \geq 1. \tag{5}$$

Using a reasoning similar to the one in [7], $\mathsf{q}_\ell$ may be expressed as a function of $\mathsf{p}_\ell$ as

$$\mathsf{q}_\ell = \mathsf{g}\left(\mathsf{p}_\ell\right) \tag{6}$$

where the function $\mathrm{g}: [0,1] \mapsto [0,1]$ is given by

$$\mathrm{g}(x) := \sum_{h=1}^{\theta} \lambda_h \frac{1}{n_h} \sum_{t=0}^{n_h-1} x^t (1-x)^{n_h-1-t} \times \\ \left[ (n_h - t) \tilde{e}^{(h)}_{n_h-t} - (t+1) \tilde{e}^{(h)}_{n_h-1-t} \right] \quad (7)$$

and where $\tilde{e}^{(h)}_t$ is the $h$-th un-normalized information function of code $C_h \in \mathcal{C}$. An expression of $\mathrm{p}_\ell$ as a function of $\mathrm{q}_{\ell-1}$ may also be obtained, following an approach similar to the one in [10]. The probability $\mathrm{p}^{(i)}_\ell$ is given by

$$\mathrm{p}^{(i)}_\ell = 1 - \sum_{k=0}^{\min(K,i)-1} \binom{i-1}{k} \mathrm{q}^k_{\ell-1} (1-\mathrm{q}_{\ell-1})^{i-k-1}. \quad (8)$$

The $k$-th summand in (8) refers to the probability that $k$ out of $i$ ESs were removed from the slice in step 2 of iteration $\ell-1$, while the sum captures the MPR capability. Next, defining

$$\rho^{(k)}(x) = \sum_{i=k+1}^{\infty} \rho_i \frac{(i-1)!}{(i-k-1)!} x^{i-k-1}, \; k > 0 \quad (9)$$

and incorporating (8) into (4), it is easy to show that

$$\mathrm{p}_\ell = 1 - \sum_{k=0}^{K-1} \frac{\mathrm{q}^k_{\ell-1}}{k!} \rho^{(k)}(1-\mathrm{q}_{\ell-1}). \quad (10)$$

The previous equation may be written in the compact form

$$\mathrm{p}_\ell = \mathrm{f}(\mathrm{q}_{\ell-1}) \quad (11)$$

where, after some manipulation, the function $\mathrm{f}: [0,1] \mapsto [0,1]$ assumes the form

$$\mathrm{f}(x) := 1 - \exp(-\mathsf{G}_0) \sum_{k=0}^{K-1} \frac{1}{k!} (x\mathsf{G}_0)^k \sum_{i=0}^{\infty} \frac{1}{i!} [(1-x)\mathsf{G}_0]^i \\ = 1 - \exp\left(-x\frac{\mathsf{G}}{\mathsf{R}}\right) \sum_{k=0}^{K-1} \frac{1}{k!} \left(x\frac{\mathsf{G}}{\mathsf{R}}\right)^k. \quad (12)$$

Note that the load $\mathsf{G}$ appears as a parameter in function $\mathrm{f}(\cdot)$. The pair of equations (6) and (11) yields a discrete dynamical system in the form $\mathrm{q}_\ell = (\mathrm{g} \circ \mathrm{f})(\mathrm{q}_{\ell-1})$ with initial state $\mathrm{q}_0 = 1$, which allows analyzing the evolution of the probability $\mathrm{q}_\ell$ in the setting $N \to \infty$ and $M \to \infty$ for constant $\mathsf{G} = N/M$.

*Definition 1 (Load threshold):* For a given CSA scheme defined by the pair $(\mathcal{C}, \Lambda)$, the supremum of the set of all $\mathsf{G} > 0$ such that, for $\ell \to \infty$, the discrete dynamical system $\mathrm{q}_\ell = (\mathrm{g} \circ \mathrm{f})(\mathrm{q}_{\ell-1})$ with initial state $\mathrm{q}_0 = 1$ converges to zero is called the load threshold and denoted by $\mathsf{G}^\star = \mathsf{G}^\star(\mathcal{C}, \Lambda, K)$.

The load threshold $\mathsf{G}^\star$ represents the largest load for which, in the limit of asymptotically large frames and with infinite number of iterations, the probability of decoding success can be made arbitrarily close to 1 with the given $(\mathcal{C}, \Lambda)$ pair over a $K$-MPR channel. For this reason, the load threshold $\mathsf{G}^\star$ is also referred to as the *capacity* of the CSA scheme. In the next section, we show that there exists a fundamental trade-off between $\mathsf{R}$ and $\mathsf{G}^\star$, as observed for collision channels in [7].

## IV. Asymptotic Performance

### A. Converse Bound on Rate-Load Pairs

Next, we develop a converse bound on the capacity of any CSA scheme of rate $\mathsf{R}$ over a $K$-MPR channel. As opposed to the load threshold $\mathsf{G}^\star$, which depends directly on $\mathcal{C}$ and $\Lambda$, the bound depends on these parameters only through the rate.

*Theorem 1:* The equation

$$\frac{\mathsf{G}}{K} = 1 - \frac{1}{K} \exp\left(-\frac{\mathsf{G}}{\mathsf{R}}\right) \sum_{k=0}^{K-1} \frac{K-k}{k!} \left(\frac{\mathsf{G}}{\mathsf{R}}\right)^k \quad (13)$$

has a unique positive solution $\mathbb{G}(\mathsf{R}, K)$ for any $0 < \mathsf{R} \leq 1$ and any integer $K \geq 1$. Moreover: (i) the CSA load threshold $\mathsf{G}^\star(\mathcal{C}, \Lambda, K)$ over the $K$-MPR channel fulfills

$$\mathsf{G}^\star(\mathcal{C}, \Lambda, K) \leq \mathbb{G}(\mathsf{R}, K) \quad (14)$$

for any rate $0 < \mathsf{R} \leq 1$, any integer $K \geq 1$, and any $(\mathcal{C}, \Lambda)$ pair corresponding to a rate $\mathsf{R}$; (ii) for any rate $0 < \mathsf{R} \leq 1$, the quantity $\mathbb{G}(\mathsf{R}, K)/K$ increases monotonically with $K$.

*Proof:* Let $A_\mathrm{g} = \int_0^1 \mathrm{g}(x) \mathrm{d}x$ and $A_\mathrm{f} = \int_0^1 \mathrm{f}(x) \mathrm{d}x$. Following a reasoning similar to the one adopted in [7] for the collision channel, if $\mathrm{q}_\ell \to 0$ as $\ell \to \infty$ then

$$A_\mathrm{g} + A_\mathrm{f} \leq 1. \quad (15)$$

Application of the Area Theorem [15] yields $A_\mathrm{g} = \mathsf{R}$ irrespective of the $(\mathcal{C}, \Lambda)$ pair. Moreover, by recursively applying integration by parts, it can be shown that

$$A_\mathrm{f} = 1 - \frac{\mathsf{R}}{\mathsf{G}} \left( K - \exp\left(-\frac{\mathsf{G}}{\mathsf{R}}\right) \sum_{k=0}^{K-1} \frac{K-k}{k!} \left(\frac{\mathsf{G}}{\mathsf{R}}\right)^k \right). \quad (16)$$

Substituting the expressions for $A_\mathrm{g}$ and $A_\mathrm{f}$ in (15) yields an inequality representing a necessary condition (*converse*) for the existence of an achievable $(\mathsf{R}, \mathsf{G})$ pair over a $K$-MPR channel. This inequality must be valid also for $\mathsf{G} = \mathsf{G}^\star$; after simple manipulation we obtain

$$\frac{\mathsf{G}^\star}{K} \leq 1 - \frac{1}{K} \exp\left(-\frac{\mathsf{G}^\star}{\mathsf{R}}\right) \sum_{k=0}^{K-1} \frac{K-k}{k!} \left(\frac{\mathsf{G}^\star}{\mathsf{R}}\right)^k. \quad (17)$$

Next, we recast (13) as $\mathsf{R} \mathrm{x} = g_K(\mathrm{x})$, where $\mathrm{x} = \mathsf{G}/(K\mathsf{R}) > 0$ and $g_K(\mathrm{x}) = 1 - \frac{1}{K} \exp(-K\mathrm{x}) \sum_{k=0}^{K-1} \frac{K-k}{k!} (K\mathrm{x})^k$. It is easy to verify that the following properties hold: $g_K(0) = 0$; $\lim_{\mathrm{x} \to +\infty} g_K(\mathrm{x}) = 1$; $g'_K(\mathrm{x}) > 0$, $\forall \mathrm{x} > 0$; $g'_K(0) = 1$; $\lim_{\mathrm{x} \to +\infty} g'_K(\mathrm{x}) = 0$; $g''_K(\mathrm{x}) < 0$, $\forall \mathrm{x} > 0$. These properties imply that the equation $\mathsf{R} \mathrm{x} = g_K(\mathrm{x})$ has a unique solution $\mathrm{x}$ for all $0 < \mathsf{R} \leq 1$. The existence and uniqueness of $\mathbb{G}(\mathsf{R}, K)$ follow from $\mathbb{G}(\mathsf{R}, K) = \mathrm{x} K \mathsf{R}$.

Claim (i) is easily proved by letting $\mathrm{x}^\star = \mathsf{G}^\star/(K\mathsf{R})$ which, with (17), yields $\mathsf{R} \mathrm{x}^\star \leq g_K(\mathrm{x}^\star)$. From the above properties of function $g_K(\mathrm{x})$, it is easy to recognize that the interval of $\mathrm{x} > 0$ in which $\mathsf{R} \mathrm{x} \leq g_K(\mathrm{x})$ corresponds to $0 \leq \mathrm{x} \leq \mathrm{x}$, where $\mathsf{R} \mathrm{x} = g_K(\mathrm{x})$ and $\mathrm{x}$ is unique as previously proved. Hence, we must have $\mathrm{x}^\star \leq \mathrm{x}$, implying $\mathsf{G}^\star(\mathcal{C}, \Lambda, K) \leq \mathbb{G}(\mathsf{R}, K)$.

To prove claim (ii), it suffices to show that $\mathrm{x} = \mathbb{G}(\mathsf{R}, K)/(K\mathsf{R})$ increases monotonically with $K$. This can be showed by defining $\Delta_K(\mathrm{x}) = g_{K+1}(\mathrm{x}) - g_K(\mathrm{x})$ and proving that $\Delta_K(\mathrm{x}) > 0$, $\forall \mathrm{x} > 0$. In this respect, it can be showed that, for some $x_1$ and $x_2$ (both depending on $K$): $\Delta_K(0) = \Delta'_K(0) = 0$; $\Delta_K(x)$ is positive, monotonically increasing and convex for $0 < x < x_1$; $\Delta_K(x)$ is positive and concave for $x_1 < x < x_2$ and has a unique maximum between $x_1$ and $x_2$; $\Delta_K(x)$ is positive, monotonically decreasing and convex for $x > x_2$; $\lim_{x \to +\infty} \Delta_K(x) = \lim_{x \to +\infty} \Delta'_K(x) = 0$. ∎

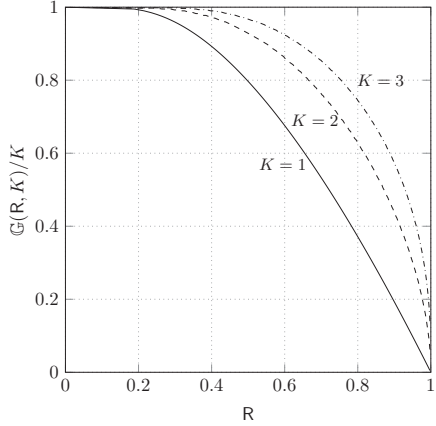

Fig. 2. Normalized converse $\mathbb{G}(\mathsf{R}, K)/K$ versus the rate for $K \in \{1, 2, 3\}$.

*Remark 1:* In the case $K = 1$ (collision channel), the bound $\mathbb{G}(\mathsf{R}, 1)$ coincides with the one developed in [7, Theorem 4.1].

Fig. 2 shows the normalized converse bound $\mathbb{G}(\mathsf{R}, K)/K$ versus the rate R, for $K \in \{1, 2, 3\}$. Coherently with Theorem 1, $\mathbb{G}(\mathsf{R}, K)/K$ *increases* with $K$ for any rate, a behavior sharply contrasting the one reported in [10] for the normalized load thresholds of IRSA schemes with a fixed maximum number of replicas. This raises the question whether this bound is meaningful, i.e., if actual SIC-based schemes exist whose normalized load thresholds $\mathsf{G}^\star/K$ approach it and, thus, benefit from MPR capabilities at the receiver. The answer to this question is positive, as exposed in the following.

### B. Approaching the Bound via Spatially-Coupled CSA

For the collision channel model, i.e., $K = 1$, it was shown that a spatially-coupled (SC) CSA tightly approaches the limit $\mathbb{G}(\mathsf{R}, K)/K$ [16]. This inspired us to extend the analysis of the regular SC-CSA schemes to the $K$-MPR channel and compare their normalized load thresholds with the developed bound.

As an example, we consider two schemes in which every user encodes its packets with a $(3, 1)$ repetition code and with a $(4, 1)$ repetition code, respectively. The rates of the schemes are $\mathsf{R} = 1/3$ and $\mathsf{R} = 1/4$, respectively; their normalized capacities are listed in Table I for $K \in \{1, 2, 3\}$ and compared with the corresponding normalized converse bound $\mathbb{G}(\mathsf{R}, K)/K$. Clearly, in both cases the normalized load threshold $\mathsf{G}^\star/K$ *increases* with $K$, although the number of packet replicas per user is constant.

A fundamental observation concerns the comparison between each actual $\mathsf{G}^\star/K$ value in Table I and the corresponding converse values. The actual values of the normalized load thresholds are all close to the bound calculated for the same R and $K$ and, notably, they approach the bound very tightly for increasing $K$. This leads us to the conclusion that the developed converse bound is indeed meaningful.

TABLE I
NORMALIZED LOAD THRESHOLDS AND CORRESPONDING CONVERSE FOR RATE 1/3 AND 1/4 SC CSA SCHEMES BASED ON REPETITION CODES

| R | | $K = 1$ | $K = 2$ | $K = 3$ |
|---|---|---|---|---|
| $\dfrac{1}{3}$ | $\mathsf{G}^\star/K$ | 0.9178 | 0.9880 | 0.9965 |
| | $\mathbb{G}(\mathsf{R}, K)/K$ | 0.9405 | 0.9895 | 0.9974 |
| $\dfrac{1}{4}$ | $\mathsf{G}^\star/K$ | 0.9767 | 0.9979 | 0.9985 |
| | $\mathbb{G}(\mathsf{R}, K)/K$ | 0.9802 | 0.9983 | 0.9988 |

## V. CONCLUSIONS

A converse bound on the normalized asymptotic load threshold of CSA with $K$-MPR capabilities at the receiver has been presented. The bound on $\mathsf{G}^\star/K$ increases with $K$, suggesting that investing in the $K$-MPR capability may pay off, even if the amount of resources required to achieve it scales linearly with $K$. The possibility to approach the bound in the asymptotic setting has been shown by means of a SC-CSA scheme.